\documentclass{aa}

\usepackage{graphicx}
\usepackage{txfonts}
\usepackage{textcomp}
\usepackage{natbib}
\bibpunct{(}{)}{;}{a}{}{,}

\begin{document}

   \title{Transmission spectroscopy of the inflated exo-Saturn HAT-P-19b\thanks{Based on observations made with the Gran Telescopio Canarias (GTC), installed in the Spanish Observatorio del Roque de los Muchachos of the Instituto de Astrofisica de Canarias, in the island of La Palma, as well as on data obtained with the STELLA robotic telescope in Tenerife, an AIP facility jointly operated by AIP and IAC.}}
   \titlerunning{Transmission Spectroscopy of HAT-P-19b}

   \author{M. Mallonn \inst{1}, C. von Essen \inst{2}, J. Weingrill \inst{1}, K.G. Strassmeier \inst{1}, I. Ribas \inst{3}, T.A. Carroll \inst{1}, E. Herrero \inst{3}, T. Granzer \inst{1}, A.~Claret \inst{4} and A. Schwope \inst{1}}
   \authorrunning{M. Mallonn et al.}

   \institute{Leibniz-Institut f\"{u}r Astrophysik Potsdam, An der Sternwarte 16, D-14482 Potsdam, Germany\\
              \email{mmallonn@aip.de}
         \and
             Stellar Astrophysics Centre, Ny Munkegade 120, 8000, Aarhus, Denmark
         \and
             Institut de Ciències de l’Espai (CSIC–IEEC), Campus UAB, Facultat de Ciències, Torre C5 parell, 2a pl, 08193 Bellaterra, Spain
         \and
             Instituto de Astrof\'{i}sica de Andaluc\'{i}a, CSIC, Apartado 3004, 18080 Granada, Spain
             }

   \date{Received; accepted}

  \abstract
   {Transiting highly inflated giant planets offer the possibility of characterizing their atmospheres. A fraction of the starlight passes through the high-altitude layers of the planetary atmosphere during transit. The resulting absorption is expected to be wavelength dependent for cloud-free atmospheres with an amplitude of up to $10^{-3}$ of the stellar flux, while a high-altitude cloud deck would cause a gray opacity.}
   {We observed the Saturn-mass and Jupiter-sized exoplanet HAT-P-19b to refine its transit parameters and ephemeris as well as to shed first light on its transmission spectrum. We monitored the host star over one year to quantify its flux variability and to correct the transmission spectrum for a slope caused by starspots. }
   {A transit of HAT-P-19b was observed spectroscopically with OSIRIS at the Gran Telescopio Canarias in January 2012. The spectra of the target and the comparison star covered the wavelength range from 5600 to 7600 \AA{}. One high-precision differential light curve was created by integrating the entire spectral flux. This white-light curve was used to derive absolute transit parameters. Furthermore, a set of light curves over wavelength was formed by a flux integration in 41 wavelength channels of 50 \AA{} width. We analyzed these spectral light curves for chromatic variations of transit depth.}
   {The transit fit of the combined white-light curve yields a refined value of the planet-to-star radius ratio of $0.1390 \pm 0.0012$ and an inclination of $88.89 \pm 0.32$ degrees. After a re-analysis of published data, we refine the orbital period to $4.0087844 \pm 0.0000015$ days. We obtain a flat transmission spectrum without significant additional absorption at any wavelength or any slope. However, our accuracy is not sufficient to significantly rule out the presence of a pressure-broadened sodium feature. Our photometric monitoring campaign allowed for an estimate of the stellar rotation period of $35.5 \pm 2.5$ days and an improved age estimate of $5.5^{+1.8}_{-1.3}$ Gyr by gyrochronology. The calculated correction of the transit depth for unocculted spots on the visible hemisphere was found to be well within the derived 1 $\sigma$ uncertainty of the white-light curve and the spectral data points of the transmission spectrum.}
   {}


   \maketitle
%

\section{Introduction}

Transiting extrasolar planets offer the opportunity of characterizing their atmospheres. During a planetary transit, parts of the starlight pass through an atmospheric ring at the planet's terminator region, picking up the signature of the planetary atmosphere. These spectral signatures are measurable with the current generation of instruments for some favorable planetary systems. The required spectroscopic precision necessary for the detection of these features depends on the planet-to-star radius ratio, the scale height of the planetary atmosphere, and the number of photons that can be collected from the host star during transit. First attempts to detect additional absorption during transit were already able to provide upper limits \citep{Rauer2000,Bundy2000,Moutou2001}. The sensitivity neccessary for a detection of an atmospheric signal was achieved for the first time by \cite{Charbonneau2002}, who detected absorption due to sodium in the atmosphere of HD209458b using HST/STIS optical observations. Today, 
 transmission spectroscopy mainly comprises two observational approaches: medium to high-resolution spectroscopy ($5000\lesssim R \lesssim 100000$), and low-resolution spectrophotometry ($R\lesssim 200$). The former one has a much higher sensitivity in detecting and resolving very narrow spectral features, whereas the latter technique is able to also reveal overall trends in the planetary spectrum that are caused by Rayleigh scattering, for example (e.g., \citealt{Pont2008,Pont2013}). 

Most of the exoplanet targets for which the current generation of instruments is sensitive enough to characterize the atmosphere are close-in gas giants. For an Earth-like extrasolar planet the atmospheric ring is too thin to detectably modify the host star's light. The close-in gas giants, known as hot Jupiters, tend to have equilibrium temperatures\footnote{The equilibrium temperature is a theoretical temperature of the planetary atmosphere assuming a radiative balance between the integrated absorbed flux received from the star and the integrated emitted flux by the planet. } of 1000 K and more. Their optical spectra might be dominated by broad atomic alkali absorption lines or, when hotter than about 1500~K, by absorption of TiO according to theoretical investigations \citep{Hubeny2003,Fortney2006,Fortney2010}. Clouds of condensed dust particles are another plausible scenario for hot Jupiters \citep{Woitke2003,Helling2008} in similarity to L-dwarfs, which are often described by 
assuming a cloud layer. \citet{Fortney2005} predicted that condensates even in very low densities in the probable atmosphere would significantly change the spectral appearance because of the slanted geometry under which the starlight passes through the planetary atmosphere.

Optical characterizations of hot Jupiters using spectrophotometry can already be found in the literature for multiple targets. These studies show a high diversity in their optical low-resolution spectral appearance. 

There are robust detections of optical alkali absorption features resolved in low spectral resolution $R\leq200$ for HD209458b \citep{Sing2008}, HAT-P-1b \citep{Nikolov2013}, XO-2 \citep{sing2012}, and WASP-31b \citep{Sing2015}. \cite{Murgas2014} presented a probable detection of Na for WASP-43b and a tentative detection of K was given by \cite{Nikolov2015}. The (low-resolution) non-detection in the atmospheres of HD189733b and WASP-12b can be explained by scattering particles blocking the atmospheric heights that
are possible for atomic sodium and potassium \citep{Pont2008,Sing2013}. The non-detection of alkali absorption in WASP-29b and HAT-P-32b can either be explained by clouds that strongly mute the spectral features or by subsolar abundances of sodium and potassium \citep{gibson2013,GibsonHATP32}. \cite{sing2011} found significant K absorption in the atmosphere of XO-2b using spectrophotometry with narrow-band tunable filters, giving a higher spectral resolution of $R\sim800$.

A spectral slope in the optical that is indicative of Rayleigh scattering by aerosols has been found for HD189733b \citep{Pont2008,Pont2013}, WASP-12b \citep{Sing2013}, WASP-6b \citep{Jordan2013}, and WASP-31b \citep{Sing2015}. In contrast, the optical spectra of HD209458b and XO-2b are best explained by a clear atmosphere model \citep{Sing2008,sing2011,sing2012}. HD209458b also features a Rayleigh slope shortward of 5000\,\AA, which is interpreted as scattering by H$_2$ molecules instead of aerosols \citep{Leca2008}. 

Indications for the presence of TiO in the terminator region was so far only found for HD29458b \citep{Desert2008} and WASP-12b; in the latter case, TiH is an alternative solution for the measurements \citep{stevenson2013}. While this molecular absorption in HD29458b still lacks an independent confirmation, a follow-up HST observation of WASP-12 found no longer any indication for TiO \citep{Sing2013}. Nor was TiO was detected in the probably cloud-free atmospheres of WASP-19b \citep{Huitson2013} and XO-2b \citep{sing2011}. Furthermore, it was ruled out for the somewhat cooler atmospheres of HAT-P-1b \citep{Nikolov2013}. The non-detection of TiO in HD189733b, HAT-P-32b, WASP-31b, and WASP-6b can be explained by clouds or haze on top of the atmosphere \citep{Pont2008,GibsonHATP32,Sing2015,Nikolov2015}.

Here we report a search for sodium absorption in the atmosphere of HAT-P-19b using ground-based differential spectrophotometry. This ground-based technique was pioneered by \citet{bean2010} for the super-Earth GJ1214b as the search for chromatic transit depth variations among a set of simultaneously observed light curves. We observed one transit event with the optical spectrograph OSIRIS at the Gran Telescopio de Canarias. Several studies already proved the capability of OSIRIS for optical transit spectrophotometry \citep{sing2011,sing2012,Colon2012,Murgas2014}.

HAT-P-19b is a hot Jupiter with a period very close to four days (\citealt{hartman2011}, hereafter H11). It is similar to Jupiter in size (1.13 $\mathrm{R}_\mathrm{J}$) and to Saturn in mass (0.29 $\mathrm{M}_\mathrm{J}$), giving it a very low surface gravity of about 6 m/s$^2$. The equilibrium temperature was estimated by H11 to be 1010 K. Hence, the scale height is $\sim600$ km assuming Jupiter's mean molecular weight of 2.2 g/mole. The planet transits a K-type main sequence star of $\mathrm{V}\,=\,12.9\,\mathrm{mag}$, $M\,=\,0.84\,\mathrm{M}_{\odot}$ and $ R\,=\,0.82\,\mathrm{R}_{\odot}$. A great advantage for differential spectrophotometry is the presence of a near-by reference star, very similar in brightness, just 1.5 arcminutes away and of similar spectral type. 

If present, brightness inhomogeneities on the stellar surface affect the derivation of transit parameters and mimic a slope in the transmission spectrum. The knowledge of a long-term flux variation and the stellar flux level at the time of the transit observation can be used to correct for this influence. For this reason, we performed a two-color monitoring campaign for HAT-P-19 using the 1.2m telescope STELLA on Tenerife over a time span of about 300 days. Similar studies can be found in the literature for the hot-Jupiter host stars HD189733 \citep{Pont2013}, WASP-19 \citep{Huitson2013}, WASP-12 \citep{Sing2013}, HAT-P-1 \citep{Nikolov2013}, WASP-6 \citep{Nikolov2015}, and WASP-31 \citep{Sing2015}.

This paper is structured as follows: Sect. 2 describes the observations and the data reduction, and Sect. 3 presents the analysis and results. A discussion of the results is given in Sect. 4, followed by the conclusions in Sect. 5.

%

\section{Observations and data reduction}

We observed the exoplanet host star HAT-P-19 in service mode during one transit event with the Gran Telescopio CANARIAS (GTC) located at the Observatorio del Roque de los Muchachos on the island of La Palma and operated by the Instituto de Astrof\'{\i}sica de Canarias. We monitored the star with the STELLA 1.2m telescope of the Leibniz-Institute for Astrophysics Potsdam (AIP) installed in the Observatorio del Teide on the island of Tenerife. 

\subsection{Spectroscopic GTC OSIRIS observation}

\subsubsection{Instrument setup}

During the transit observation we made use of GTC's optical spectrograph OSIRIS. The dispersive element was the VPH R2500R, providing a wavelength coverage of ~5600 to 7500 \AA{} and a dispersion of about 1 \AA{} per binned pixel. We used the widest long slit available at the time of the observation, the $5.0''$ slit, which caused the spectral resolution to be seeing limited to roughly $R\sim1000$. We read out the two Marconi CCD detectors in the fastest available, 500~kHz read-out full-frame mode. The standard 2x2 binning resulted in less than 10 seconds closed-shutter time between the exposures. Exposure time was 60 seconds. 

We chose a rotation angle of the instrument that centered (in dispersion direction) both the target HAT-P-19 and the reference star GSC 0228301197 (USNO-B1 1246-0009792) in the slit for simultaneous observations. OSIRIS covers its field of view with two CCD chips; we placed both objects on chip 1 to avoid potential differences between chip 1 and chip 2 as error sources in the differential spectrophotometry. 

\subsubsection{Observing log}

The transit observation was conducted on January 10, 2012. The target was observed for slightly less than five hours from 19:33 to 00:19 UT, resulting in 238 exposures. We typically reached a S/N ratio of $\sim$380 per pixel in dispersion at central wavelengths. The transit lasted from 19:56 to 22:39 UT (first to fourth contact). The night was photometric with the seeing varying from $0.9''$ to $1.3''$. Nevertheless, the observing conditions were far from optimal for high-precision spectrophotometry. The observation began shortly after the meridian passage of HAT-P-19, which occurred very close to the zenith. Unfortunately, the dome construction of the GTC could not be fully opened at the time of the observation, causing vignetting at the highest elevations. This vignetting was different for target and comparison star, clearly distracting the transit light-curve (Fig. \ref{wlc}) during pre-transit phase. Moreover, the target set quickly during the time series to a maximum airmass of 3.23, leading to a rapid 
spread of the point spread function with time. In Fig. \ref{plot1} the evolution of the flux level, airmass, FWHM, and detector position relative to its mean is shown. The point spread function became asymmetric at about half the observing time due to a loss of focus; the observations were not stopped for focus adjustment to avoid further systematics introduced by an abrupt change of the telescope settings. Therefore the obtained values for the FWHM and pixel position need to be treated with caution because they were derived by fitting a Gaussian function to the spatial spectral profile. However, the guiding system kept the spectrum almost at the same position during the entire time series, it moved in spatial direction by less than 1 pixel. Two jumps in pixel position occurred, which were of subpixel value and had no effect on the resulting light curves. 

\subsubsection{Data reduction}

The OSIRIS data were reduced by routines written in ESO-Midas. Standard calibration frames have been obtained at the same day as the science observations. Flatfield frames were exposed through the 5$''$ wide long slit, the same slit as used for the science frames. However, calibration spectra of HgAr and Ne lambs were taken with a 1$''$ wide slit to avoid unpractically broad emission lines. A bias subtraction of the flat, arc lamp, and science frames was performed using the overscan regions. The flatfield correction was done by a master flat created of the flux-weighted average of 101 single flatfield frames taken the same day. We noticed that the flatfield correction had almost no effect on the quality in terms of scatter on the final transit light-curves or on the absolute value of the transit parameters. For example, the difference in the planet-to-star radius ratio $k$ whether a flat field correction was applied or not was about ten percent of the derived 1 $\sigma$ uncertainty.

Before the wavelength calibration we corrected for a drift of the spectra in dispersion direction of about 1.4 pixel over the length of the time series. The origin of this drift is most probably a combination of telescope flexure, obvious in a shift of the skylines by a total of about 0.4 pixel, and a drift of the chromatic centroids of the objects in the long slit, probably caused by field rotation residuals and differential atmospheric refraction (OSIRIS does not include an atmospheric dispersion corrector). For simplicity we corrected for the drift achromatically. A difference in drift among the two objects has just been found for very high airmass, data that were excluded in the later analysis (see Sect. \ref{wlc_ana}), hence we applied the same drift correction over the entire spatial range. In the through-slit acquisition image, taken in the Sloan r band, we measured a slight mismatch of the object centroids in dispersion of $\sim$0.4 pixel ($\sim$0.1$''$), which we also 
corrected for after the wavelength calibration and spectral extraction. In summary, we estimate the wavelength calibration to be accurate to just about 0.5 \AA{} due to residual wavelength dependent shifts and stresses over time both in dispersion and spatial direction.

In the presence of curved sky emission lines, we decided for a two-dimensional wavelength calibration of the whole frame to allow for sky estimation in spatially extended stripes distant to the spatial center of the object. While the choice of the sky stripe width minimized the light-curve scatter by up to ten percent depending on the wavelength region, it influenced the absolute transit parameters always by less than their 1 $\sigma$ error values and did not affect the derived transmission spectrum either. The final width of the sky stripes spatially above and below the spectrum was 100 pixel each. The stripes were separated from the object extraction stripe by a 50 pixel gap due to the very extended wings of the spectral profile. The sky value at the object position was estimated by a linear interpolation between these two sky stripes, independently derived for each pixel in dispersion direction. 

\begin{figure}
   \centering
   \includegraphics[width=\hsize]{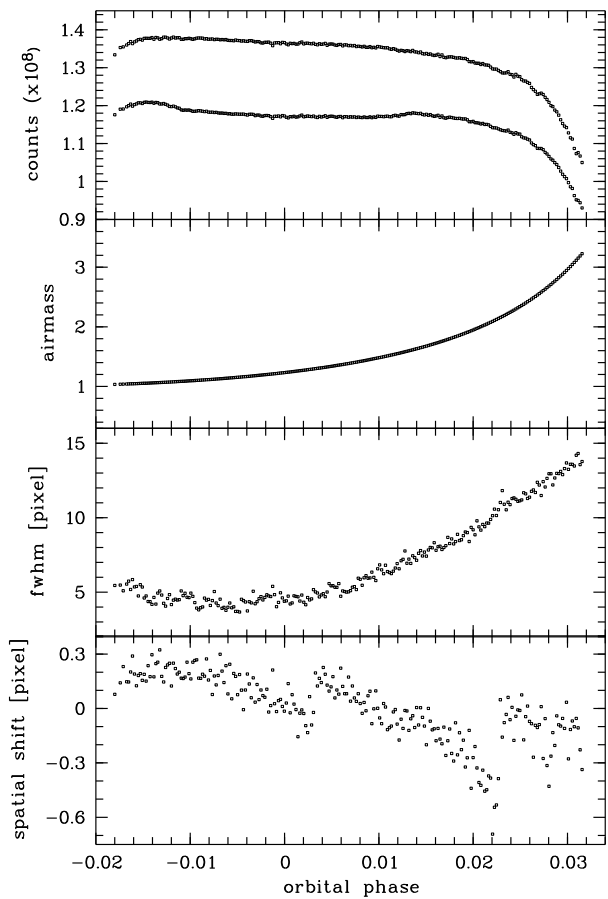}
      \caption{Evolution of observational parameters over the time series. From top to bottom: count rate of target (lower flux level) and reference star (higher flux level) in ADU integrated over the spectral range, airmass, FWHM in spatial direction at central wavelength in pixel, and spectrum displacement in spatial direction in pixel.}
         \label{plot1}
   \end{figure}

\begin{figure}
   \centering
   \includegraphics[height=\hsize,angle=270]{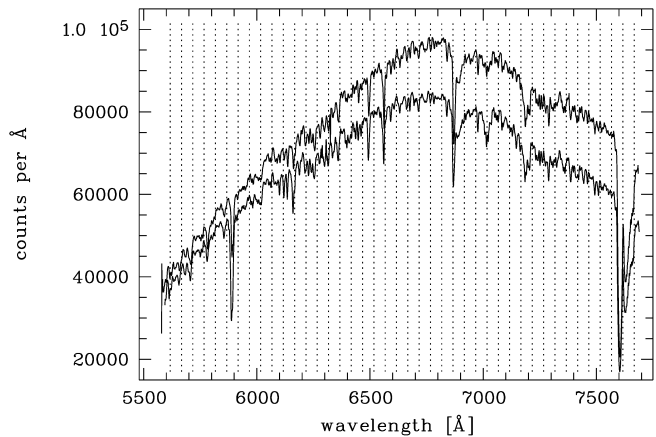}
      \caption{Spectra of HAT-P-19 (lower flux level) and the comparison star (higher flux level) of the first exposure. The 50 \AA{} wavelength channels used for spectrophotometry are indicated by vertical dotted lines.}
         \label{exspec}
   \end{figure}

The widths of the aperture in which to extract the object flux were chosen to scale with the spatial FWHM of the spectral profile to account for the immense expansion of the point spread function at higher airmass (Fig. \ref{plot1}). A set of scaling factors were tested to minimize the scatter in the white-light curve. The final aperture size was 7.5 times the FWHM, which ranges from 25 to 105 pixel within the time series. Different widths of the sky stripes  were alos tested. The scatter of the white-light curve decreased slowly with increasing width until it reached a plateau at about 100 pixels. We determined whether the optimal extraction technique of \citet{Horne86} could further minimize the dispersion on the light curves, but as expected for very high S/N spectra, there was no significant difference to a simple flux sum in spatial direction. Instead, the simple flux sum yielded more robust results in the few regions with hot or dark pixels. The one-dimensional spectra were extracted in pieces of steps 
10 \AA{} wide 
 to account for the wavelength-dependent FWHM of the spectral profile. Then the pieces were stitched together in wavelength, forming spectra from 5617 to 7687 \AA{}. The spectra of exposure 1 are shown in Fig. \ref{exspec}.

The spectra were then divided into wavelength channels of a certain width (see Fig. \ref{exspec} and Sect. \ref{chomatic_lcs}). For each channel the flux of both objects, target and comparison star, was integrated and used to perform differential photometry. In this way, a set of simultaneously observed light curves consecutive in wavelength was achieved. Individual photometric errors were estimated by
\begin{equation}
\Delta F \, = \, \frac{\sqrt{A\sigma^2 + \frac{F}{g}}}{F}\, ,
\end{equation}

with $A$ as the area (in pixel) over which the total flux $F$ (in ADU) was summed, $\sigma$ the standard deviation of noise (in ADU) estimated in the two sky stripe areas, and $g$ the detector gain of $g=1.46\,\,e^-$/ADU. This equation follows the photometric error estimation of the software package Source Extractor\footnote{www.astromatic.net/software/sextractor} \citep{Bertin96}.

\subsection{Photometric STELLA-WIFSIP monitoring}

\subsubsection{Instrument setup}

We observed HAT-P-19 with the wide-field imager WIFSIP of the robotic 1.2m twin-telescope STELLA on Tenerife \citep{Strassmeier2004,Weber2012}. WIFSIP consists of a 4kx4k back-illuminated CCD with a plate scale of 0.322 ``/pixel and four read out amplifiers. It covers a field of 22 by 22 arcminutes on the sky. The automatic scheduler of STELLA was set to observe HAT-P-19 on average every second night in blocks of three exposures in V and three exposures in I of 20 seconds exposure time each. We obtained 324 frames within 35 different observing nights from December 2011 to February 2012; between May 2012 and October 2012, the dataset contains 1268 frames from 82 nights. 

\subsubsection{Data reduction}

The bias and flatfield correction was made with the STELLA data reduction pipeline, for details see Granzer et al. (2015, in preparation). The following reduction steps were made with routines written in ESO-MIDAS. We conducted aperture photometry with the software package Source Extractor. Its MAG\_AUTO option calculates an elliptic aperture individually for each image and object according to the second-order moments of the object's light distribution. This method provides the flexibility to account for the varying observing conditions over the ten-month observing time. The I-band data suffered from fringing, whose pattern was found to depend mainly on CCD detector temperature. We created master fringe maps by averaging the individual object-removed and smoothed science frames in groups of the same CCD temperature. The fringe correction was made by subtracting the temperature-selected master fringe map, scaled in amplitude to match the fringes of the individual science frames. The 
fringe 
residuals typically had a strength of about 20 percent of the original fringe pattern.    

We used the method of optimal weighting of an ensemble of comparison stars \citep{Broeg05} to form an artificial reference for differential photometry. We experimented with many different comparison star ensembles to verify that the observed photometric signal was independent of the choice of the comparison stars. The final ensemble is formed by three stars similar in brightness and color to HAT-P-19, read out through the same amplifier as the target. The same ensemble was used for the V- and the I-band data. After averaging the three consecutively taken data points and subtracting a transit model using the transit parameter derived in Sects. \ref{wlc_ana} and \ref{chap_ephe} , the point-to-point scatter of the data versus a sine model (see Sect. \ref{ana_lotephot}) was about 2.5~mmag in both filters.

\section{Analysis and results}

We fit \cite{Mandel2002} transit light-curve models to all transit light-curves. The main parameters of the fit are the inclination $i$, the transit midpoint $T_0$, the limb-darkening coefficients of a quadratic law $u_{\mathrm{A}}$ and $v_{\mathrm{A}}$, the scaled semimajor axis $a / R_A $ , and the ratio of the planetary and stellar radius $ k\,=\,R_b/R_A$, where $a$ is the semimajor axis of the planetary orbit and $R_b$ and $R_A$ are the absolute planetary and stellar radii. The best-fitting parameters were determined by a least-squares Markov chain Monte Carlo (MCMC) approach that made use of PyAstronomy\footnote{https://github.com/sczesla/PyAstronomy}.

\subsection{Detrending and white-light curve analysis}
\label{wlc_ana}

\begin{figure}
   \centering
   \includegraphics[height=\hsize,angle=270]{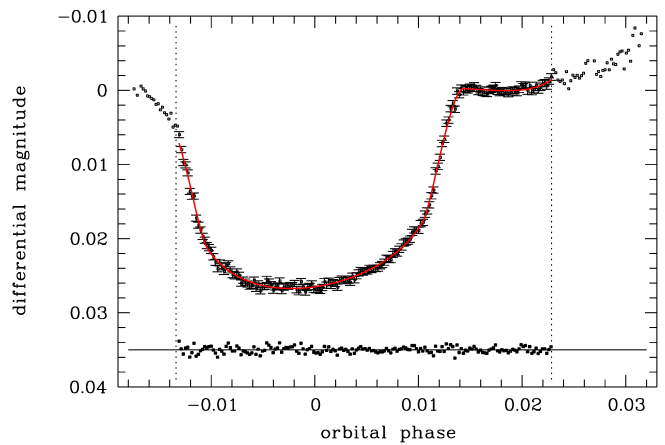}
      \caption{Transit light-curve integrated over the entire spectral range (white light). The vertical dashed lines confine the data that were used for the analysis, see text for details. The red solid line shows the best-fit model, the residuals are presented at the bottom.}
         \label{wlc}
   \end{figure}

A white-light curve was formed by integrating the flux of the entire spectral range. As a very first step, we excluded the first 20 data points that lost flux because of the obscuration of the primary mirror by the dome construction. Unfortunately, these data formed our pre-ingress measurements. The resulting light-curve features obviously smooth deviations from a theoretical symmetric transit light-curve, see Fig. \ref{wlc}. To model this deformation, we fitted a low-order polynomial to external parameters simultaneously to the transit model. Various polynomials to the external parameters time, airmass $z$, sky background, FWHM in spatial direction, pixel shift in x (dispersion direction), and pixel shift in y (spatial direction), and their combinations were tested. The pixel shift in spatial direction was measured by the fit of a Gaussian function to the spatial profile, the measurement of the pixel shift in dispersion was taken as the average of the centroid of six spectral features per 
object. 

The fit parameters of the transit model were $a / R_A$, $ k $ , $i$, and $T_0$ with the values determined by H11 as starting values plus their derived errors as Gaussian priors. The eccentricity and longitude of periastron were fixed to the values of H11. We ran ten chains of $1 \times 10^5$ MCMC iterations, disgarded the initial $3 \times 10^4$ steps in every chain as burn-in phase, and merged all remaining iterations. The best-fit parameter values were derived as the mean of the posterior parameter distribution and the 1 $\sigma$ uncertainties as the 68.3\% of highest probability.
 
To avoid overfitting of the deformations andtrends in the light curve, we chose the function that minimized the Bayesian information criterion (BIC, \citealt{Schwarz78}). The BIC is similar to the goodness-of-fit estimation with a chi-square calculation, but adds a penalty term for the number of parameters in the model,
 \begin{equation}
BIC\,=\, \sum \dfrac{(O - M)^2}{\sigma^2} \,+\,m \cdot \mathrm{ln}(n)
,\end{equation}
with $O$ in our case as the observation, $M$ as the transit and detrend model, $m$ the number of free parameters, $\sigma$ the photometric error of the individual data points, and $n$ the number of data points, here 175. We expect Eq. 1 to underestimate these values because of its ignorance of correlated noise. Therefore, we calculated a scaling factor to match the mean photometric error with the standard deviation of the photometric points. To estimate this factor, we only used the post-transit data (phase > 0.015) to avoid any bias introduced by a premature transit modeling. Then the BIC values for the tested low-order polynomials were calculated on the entire time series. Table \ref{tabBIC} gives these values for a selection of tested detrending functions together with their derived transit parameters, which depend significantly on the detrending function choice. The final choice for the detrending function was a third-order polynomial with the 
airmass as independent variable because it minimized the BIC value, the scatter of the residuals (rms), and the amount of residual time-correlated noise (estimated by the $\beta$ value, see below).   

\onltab{
\begin{table*}
\tiny
\caption{Dependence of the derived transit parameters on the model function used for detrending. The notation $p(j^i)$ refers to a polynomial of degree i of parameter j, e.g., $p(z^3)$ denotes a polynomial of third degree with respect to airmass $z$. The model function used for the final light curve analysis is printed in boldface.}
\label{tabBIC}
\begin{center}
\begin{tabular}{ccccccc}

\hline
model    & BIC & rms    & $\beta$ & $k=R_b/R_A$ & $ a/R_A $ & $i$  \\
function &     & (mmag) &         &             &               & (deg)\\
\hline
\hline
\noalign{\smallskip}
$p(z^4)$          &  292.1  &  0.42 & 1.08  & 0.1406 $\pm$ 0.0012 & 12.32 $\pm$ 0.21 &  89.20 $\pm$ 0.31 \\
\boldmath{$p(z^3)$} &  \textbf{279.8}  &  \textbf{0.43} & \textbf{1.12}  & \textbf{0.1390 $\pm$ 0.0012}  &  \textbf{12.37 $\pm$ 0.21} &  \textbf{88.89 $\pm$ 0.32} \\
$p(z^2)$          &  328.7  &  0.48 & 1.58  & 0.1341 $\pm$ 0.0020 & 12.41 $\pm$ 0.43 &  88.95 $\pm$ 0.61 \\
$p(z^3) + p(x^1)$ &  292.2  &  0.43 & 1.17  & 0.1379 $\pm$ 0.0012 & 12.46 $\pm$ 0.25 &  89.12 $\pm$ 0.45 \\
$p(z^3) + p(y^1)$ &  294.4  &  0.43 & 1.21  & 0.1382 $\pm$ 0.0013 & 12.23 $\pm$ 0.29 &  88.95 $\pm$ 0.43 \\
$p(z^2) + p(t^2)$ &  339.7  &  0.48 & 1.56  & 0.1343 $\pm$ 0.0015 & 12.41 $\pm$ 0.33 &  89.20 $\pm$ 0.56 \\
$p(t^4)$          &  304.3  &  0.45 & 1.09  & 0.1399 $\pm$ 0.0014 & 12.24 $\pm$ 0.27 &  88.97 $\pm$ 0.46 \\
$p(t^3)$          &  319.1  &  0.48 & 1.36  & 0.1335 $\pm$ 0.0018 & 12.56 $\pm$ 0.44 &  89.19 $\pm$ 0.64 \\
$p(t^2)$          &  316.9  &  0.50 & 1.51  & 0.1354 $\pm$ 0.0021 & 12.51 $\pm$ 0.36 &  88.85 $\pm$ 0.59 \\
$p(FWHM^4)$       &  337.4  &  0.47 & 1.71  & 0.1341 $\pm$ 0.0019 & 12.46 $\pm$ 0.43 &  88.61 $\pm$ 0.78 \\
$p(FWHM^3)$       &  393.0  &  0.54 & 1.99  & 0.1329 $\pm$ 0.0027 & 12.38 $\pm$ 0.51 &  88.95 $\pm$ 0.95 \\
                                                     
\hline
\end{tabular}
\end{center}
\end{table*}
}

The stellar limb-darkening was accounted for by the quadratic law in the transit fit. The limb-darkening coefficients were spherically calculated for a star of $T_\mathit{eff}=5000$ K and log $g=4.5$ in high spectral resolution using PHOENIX stellar atmosphere models of metallicity log[M/H]~=~0.0, see \cite{Claret13} for details. The applied values for the white-light curve were found by a flux-weighted average of these nearly monochromatic values. To analyze the white-light curve, we fitted for the linear limb-darkening coefficient $u_{\mathrm{A}}$ and fixed the nonlinear term $v_{\mathrm{A}}$ to the theoretical value. The fitted value of the linear term agreed with the prediction to within about 1 $\sigma$. This approach of fitting $u_{\mathrm{A}}$ but fixing $v_{\mathrm{A}}$ to its theoretical value 
allows for some flexibility to account for potential systematic differences between the star and its theoretical model spectrum used for determination of the limb-darkening parameters. Furthermore, it avoids the problem of enlarged uncertainties when fitting for both coefficients caused by the strong correlation between $u_{\mathrm{A}}$ and $v_{\mathrm{A}}$ \citep{Southworth07,Southworth08,Johnson2008}.

We briefly comment on the systematic error introduced by the metallicity difference between the PHOENIX models (solar metallicity) and HAT-P-19: H11 derived a value of log[M/H]~=~$0.23 \pm 0.08$. We used the limb-darkening tables of \cite{Claret2000} and \cite{Claret2004} to estimate the systematic error to be on the order of 0.005 for the quadratic limb-darkening coefficient (the linear coefficient is a free parameter in our transit fit). The resulting systematic error on the derived transit parameters is much smaller than their uncertainties.

The individual photometric error bars were enlarged by a common factor to give a reduced chi-square value of unity, $\chi_{\nu}^2=1.0 $, versus the fitted model. To account for correlated red noise, we derived in a second step an additional scaling factor as the ratio of the calculated and the theoretical standard deviation of the binned photometric residuals, often called the $\beta $ factor \citep{Gillon06,Winn08}. We binned the residuals in intervals from 10 to 20 minutes (about the duration of ingress) in one-minute steps (original time sampling is $\sim$1.1 minute), derived the factor in each case, and finally used their average as the $\beta $ factor to enlarge the individual photometric errors.

There are features in the noise in the post-transit light curve at phases later than about 0.023 visible to the eye that are believed to be systematics potentially caused by differential slit loss. For white noise alone, the uncertainty of the radius ratio $k$ should \textit{\textup{decrease}} approximately proportionally to the square root of the number of out-of-transit (OoT) data. Here it was worth trying whether the uncertainties actually decrease with \textit{\textup{fewer}} OoT data. And indeed, the uncertainty of the radius ratio $\Delta k $ reached a minimum at an OoT limit of phase 0.0228. In addition, the uncertainties of the scaled semimajor axis $\Delta (a/R_A)$, the inclination $\Delta i,$ and the transit mid-time $\Delta T_0$ reached their minima at about this number of OoT data, therefore we discarded all later phases from the analysis.

The results of the white-light curve analysis roughly agree with the work of H11 and are given in Table \ref{KapSou} together with an estimate of the surface gravity of the planet $g_b$, calculated with Eq. 4 of \cite{Southworth2007} using the value of the stellar velocity amplitude $K$ from H11.

\begin{table}
\caption{Transit fit parameter of the white-light curve. }
\label{KapSou}
\begin{center}
\begin{tabular}{ll}
\hline
\noalign{\smallskip}
Parameter      &  Value  \\
\noalign{\smallskip}
\hline
\noalign{\smallskip}
$a/R_A$   & 12.37 $\pm$ 0.21  \\
$k$              & 0.1390 $\pm$ 0.0012  \\
$i$ [deg]             & 88.89 $\pm$ 0.32   \\
$T_0$ - 2400000 [days] & 55937.38839 $\pm$ 0.00011 \\
$u_{\mathrm{A}}$             & 0.5736 $\pm$ 0.0182   \\
$v_{\mathrm{A}}$ (fixed)     & 0.1376  \\
$g_b$ [ms$^{-2}$] & 5.97 $\pm$ 0.61 \\
\noalign{\smallskip}
\hline
\end{tabular}
\end{center}
\end{table}

\onltab{
\begin{table*}
\tiny
\caption{Characteristics of the extracted transit light-curves. }
\label{rmslc1}
\begin{center}
\begin{tabular}{rrrlll}

\hline
Wavelength & $\beta$ & rms & $u_{\mathrm{A}}$ & $v_{\mathrm{A}}$ & $k=r_b/r_A$  \\
(\AA{}) &  & (mmag) &  fitted & calculated &\\
\hline
\hline
\noalign{\smallskip}
 & & & & & \\
5617 - 7616 & 1.12  & 0.43 & 0.5736 &  0.1376 & 0.1390 $\pm$ 0.0012\\
\hline
\noalign{\smallskip}
5617 - 5816 & 1.16  & 0.73 & 0.6087 & 0.0988  & 0.1412 $\pm$ 0.0013 \\
5817 - 6016 & 1.11  & 0.69 & 0.6127 & 0.1132  & 0.1382 $\pm$ 0.0015 \\
6017 - 6216 & 1.27  & 0.62 & 0.6233 & 0.1230  & 0.1390 $\pm$ 0.0014 \\
6217 - 6416 & 1.04  & 0.68 & 0.5964 & 0.1258  & 0.1392 $\pm$ 0.0014 \\
6417 - 6616 & 1.02  & 0.71 & 0.5663 & 0.1450  & 0.1392 $\pm$ 0.0014 \\
6617 - 6816 & 1.15  & 0.64 & 0.5708 & 0.1442  & 0.1383 $\pm$ 0.0012 \\
6817 - 7016 & 1.19  & 0.68 & 0.5726 & 0.1495  & 0.1388 $\pm$ 0.0010 \\
7017 - 7216 & 1.00  & 0.66 & 0.5586 & 0.1443  & 0.1392 $\pm$ 0.0012 \\
7217 - 7416 & 1.00  & 0.58 & 0.5155 & 0.1529  & 0.1407 $\pm$ 0.0011 \\
7417 - 7616 & 1.06  & 0.59 & 0.5010 & 0.1570  & 0.1412 $\pm$ 0.0012 \\
\hline
\noalign{\smallskip}
5617 - 5666 & 1.33  & 1.21 & 0.6021 & 0.0989  & 0.1422 $\pm$ 0.0027 \\
5667 - 5716 & 1.05  & 1.10 & 0.5722 & 0.0969  & 0.1423 $\pm$ 0.0023 \\
5717 - 5766 & 1.27  & 1.09 & 0.5748 & 0.0921  & 0.1408 $\pm$ 0.0021 \\
5767 - 5816 & 1.02  & 0.99 & 0.6060 & 0.1068  & 0.1380 $\pm$ 0.0017 \\
5817 - 5866 & 1.00  & 0.94 & 0.6293 & 0.1204  & 0.1358 $\pm$ 0.0017 \\
5867 - 5916 & 1.08  & 1.02 & 0.6034 & 0.0930  & 0.1394 $\pm$ 0.0018 \\
5917 - 5966 & 1.00  & 1.06 & 0.6470 & 0.1118  & 0.1359 $\pm$ 0.0017 \\
5967 - 6016 & 1.03  & 1.03 & 0.5886 & 0.1246  & 0.1379 $\pm$ 0.0019 \\
6017 - 6066 & 1.11  & 1.04 & 0.6163 & 0.1278  & 0.1377 $\pm$ 0.0026 \\
6067 - 6116 & 1.22  & 1.09 & 0.6390 & 0.1158  & 0.1345 $\pm$ 0.0020 \\
6117 - 6166 & 1.20  & 1.21 & 0.6026 & 0.1221  & 0.1414 $\pm$ 0.0022 \\
6167 - 6216 & 1.01  & 1.03 & 0.5902 & 0.1263  & 0.1389 $\pm$ 0.0017 \\
6217 - 6266 & 1.00  & 0.91 & 0.5626 & 0.1246  & 0.1397 $\pm$ 0.0015 \\
6267 - 6316 & 1.18  & 0.95 & 0.5931 & 0.1248  & 0.1384 $\pm$ 0.0016 \\
6317 - 6366 & 1.13  & 1.05 & 0.5843 & 0.1307  & 0.1378 $\pm$ 0.0025 \\
6367 - 6416 & 1.00  & 0.99 & 0.5779 & 0.1232  & 0.1393 $\pm$ 0.0015 \\
6417 - 6466 & 1.04  & 0.90 & 0.5620 & 0.1225  & 0.1401 $\pm$ 0.0018 \\
6467 - 6516 & 1.04  & 0.93 & 0.5867 & 0.1352  & 0.1381 $\pm$ 0.0019 \\
6517 - 6566 & 1.41  & 0.92 & 0.5425 & 0.1703  & 0.1402 $\pm$ 0.0016 \\
6567 - 6616 & 1.08  & 1.06 & 0.5467 & 0.1510  & 0.1379 $\pm$ 0.0021 \\
6617 - 6666 & 1.26  & 1.02 & 0.5639 & 0.1415  & 0.1391 $\pm$ 0.0018 \\
6667 - 6716 & 1.40  & 0.95 & 0.5864 & 0.1419  & 0.1364 $\pm$ 0.0016 \\
6717 - 6766 & 1.31  & 0.87 & 0.5655 & 0.1483  & 0.1386 $\pm$ 0.0016 \\
6767 - 6816 & 1.06  & 0.82 & 0.5979 & 0.1450  & 0.1383 $\pm$ 0.0015 \\
6817 - 6866 & 1.00  & 0.93 & 0.5371 & 0.1472  & 0.1394 $\pm$ 0.0018 \\
6867 - 6916 & 1.43  & 0.95 & 0.5530 & 0.1533  & 0.1390 $\pm$ 0.0014 \\
6917 - 6966 & 1.00  & 1.05 & 0.6084 & 0.1484  & 0.1375 $\pm$ 0.0017 \\
6967 - 7016 & 1.15  & 1.00 & 0.5354 & 0.1492  & 0.1391 $\pm$ 0.0017 \\
7017 - 7066 & 1.05  & 1.36 & 0.4911 & 0.1421  & 0.1369 $\pm$ 0.0029 \\
7067 - 7116 & 1.01  & 0.95 & 0.5573 & 0.1498  & 0.1392 $\pm$ 0.0014 \\
7117 - 7166 & 1.05  & 1.13 & 0.5190 & 0.1434  & 0.1393 $\pm$ 0.0024 \\
7167 - 7216 & 1.02  & 1.39 & 0.5635 & 0.1414  & 0.1385 $\pm$ 0.0022 \\
7217 - 7266 & 1.01  & 0.97 & 0.5161 & 0.1494  & 0.1404 $\pm$ 0.0014 \\
7267 - 7316 & 1.00  & 0.82 & 0.4732 & 0.1568  & 0.1418 $\pm$ 0.0016 \\
7317 - 7366 & 1.00  & 0.93 & 0.4958 & 0.1527  & 0.1403 $\pm$ 0.0016 \\
7367 - 7416 & 1.04  & 0.96 & 0.5532 & 0.1530  & 0.1382 $\pm$ 0.0016 \\
7417 - 7466 & 1.00  & 0.92 & 0.4955 & 0.1578  & 0.1406 $\pm$ 0.0017 \\
7467 - 7516 & 1.07  & 0.81 & 0.5077 & 0.1572  & 0.1399 $\pm$ 0.0015 \\
7517 - 7566 & 1.00  & 0.87 & 0.4793 & 0.1602  & 0.1394 $\pm$ 0.0015 \\
7567 - 7616 & 1.00  & 0.98 & 0.5008 & 0.1510  & 0.1432 $\pm$ 0.0018 \\
7617 - 7666 & 1.13  & 1.03 & 0.5131 & 0.1543  & 0.1394 $\pm$ 0.0018 \\
\hline
\noalign{\smallskip}
5667 - 5866 & 1.13  & 0.65 & 0.6002 & 0.1046  & 0.1395 $\pm$ 0.0013 \\
5867 - 5916 & 1.08  & 1.02 & 0.6034 & 0.0930  & 0.1394 $\pm$ 0.0018 \\
5917 - 6516 & 1.25  & 0.56 & 0.5908 & 0.1243  & 0.1386 $\pm$ 0.0012 \\
6537 - 6587 & 1.31  & 1.05 & 0.5449 & 0.1727  & 0.1397 $\pm$ 0.0020 \\
6587 - 7666 & 1.14  & 0.46 & 0.5296 & 0.1492  & 0.1391 $\pm$ 0.0011 \\
\hline
\end{tabular}
\end{center}
\end{table*}
}

\subsection{Ephemeris}
\label{chap_ephe}
\begin{figure}
   \centering
   \includegraphics[height=\hsize,angle=270]{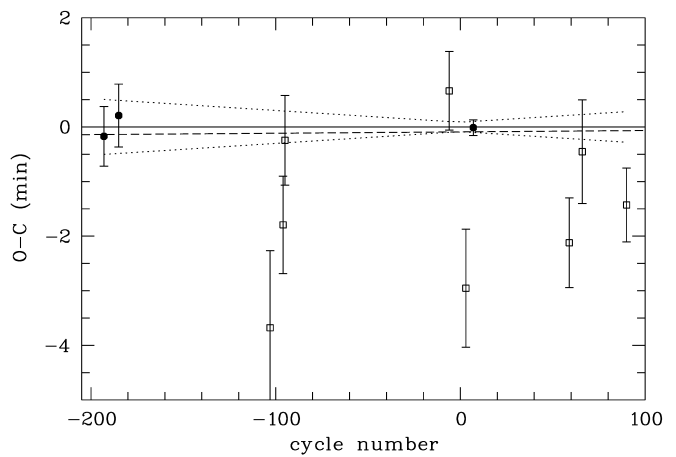}
      \caption{Transit timing residuals versus the linear ephemeris of Eq. \ref{equ_ephem}. The filled symbols represent the two transits of H11 and the transit of this work, which were used to derive the ephemeris. The open symbols show the amateur ETD transits. The dotted lines give the 1$\sigma$ uncertainty of the ephemeris as a function of cycle number. The dashed line shows the difference when the amateur data are included in the ephemeris fit, which yielded no significant improvement and was finally not used for Eq. \ref{equ_ephem}. }
         \label{plot_ephe}
   \end{figure}

We used the white-light curve of our transit measurement and the publicly available two complete transits from H11, which
were observed with the KeplerCam camera on the FLWO 1.2m telescope, to recalculate the ephemeris. The FLWO transits were reanalyzed with our algorithms to guarantee consistency in the analysis. A first-order polynomial over time was used for detrending. The free transit parameter was the mid-time of the transit $T_0$. The parameters $a / R_A$, $ k $ , and $i$ were fixed to the values derived from our white-light curve, given in Table \ref{KapSou}. The quadratic limb-darkening coefficients were fixed to the values used by H11. The results of the mid-transit times are summarized in Table \ref{tabmidtime} and graphically presented in Fig. \ref{plot_ephe}. We experimented with the coefficients, using the linear coefficient as free fit parameter or varying both in meaningful ranges. The tests showed very little effect of very few seconds on $T_0$, which is well inside 
the errors. 
H11 used the Coordinated Universal Time (UTC) as time base for their BJD calculations. As recommended by \cite{eastman2010}, we converted all times to the Barycentric Dynamical Time (TDB) using their available online tools\footnote{http://astroutils.astronomy.ohio-state.edu/time/}. The newly found ephemeris is
\begin{equation}
T_c\,=\,\mathrm{BJD(TDB)}\,2455909.326808(55)\,+\,4.0087844(15)\,N .
\label{equ_ephem}
\end{equation}
$T_c$ is the predicted central time of a transit, $N$ is the cycle number with respect to the reference mid-time, and the numbers in brackets give the uncertainties of the last two digits. The reference mid-time was chosen to minimize the covariance between reference mid-time and period. The GTC transit accordingly
corresponds to cycle  7.

We tested whether additional transit light-curves observed by amateur astronomers could further improve the accuracy of the ephemeris. The data were made publicly available on the Exoplanet Transit Database (ETD) website\footnote{http://var2.astro.cz/ETD/}. We selected the ETD transits to have uncertainties on the mid-transit time $T_0$ of about one minute or less. The light curves were reanalyzed and the transit mid-times converted into BJD(TDB). However, we found that the time uncertainties of the eight ETD transits are too large to significantly improve the ephemeris of Eq. \ref{equ_ephem}. Therefore we decided to use the better-documented and peer-reviewed observations of H11 and our own observation. The timings of ETD roughly agree with Eq. \ref{equ_ephem}, see Fig. \ref{plot_ephe}, their residuals are all consistent with zero within less than 3 $\sigma$.

\begin{table}
\tiny
\caption{Transit mid-times of HAT-P-19 and their residuals versus the ephemeris derived in this work. The upper part lists the
values of the professional light curves that were used to derive the ephemeris. The lower part lists the values of the amateur light curves (ETD/TRESCA) that were tested in the analysis, but were not used in the end to derive the ephemeris. }
\label{tabmidtime}
\begin{center}
\begin{tabular}{lrrl}

\hline
Mid-time $T_0$ & Cycle & Residual & Reference \\
(BJD(TDB) - 24000000) & number  & (days) &  \\
\hline
\hline
\noalign{\smallskip}
55135.63128 $\pm$ 0.00038 & -193  &  -0.00011 &     \cite{hartman2011} \\
55167.70182 $\pm$ 0.00040 & -185  &   0.00014 &     \cite{hartman2011} \\
55937.38829 $\pm$ 0.00010 & 7     &  -0.00001 &     This work  \\
\hline
\noalign{\smallskip}
55496.41945 $\pm$ 0.00075 & -103  &  -0.00255 &     Naves, R \\
55524.48225 $\pm$ 0.00062 & -96   &  -0.00124 &     Muler, G.  \\
55528.49211 $\pm$ 0.00057 & -95   &  -0.00017 &     Ruiz, J. \\
55885.27456 $\pm$ 0.00050 & -6    &   0.00045 &     Ayiomamitis, A.  \\
55921.35111 $\pm$ 0.00075 & 3     &  -0.00205 &     Naves, R.  \\
56145.84362 $\pm$ 0.00051 & 59    &  -0.00147 &     Shadic, S.  \\
56173.90627 $\pm$ 0.00066 & 66    &  -0.00031 &     Garlitz, J. \\
56270.11642 $\pm$ 0.00047 & 90    &  -0.00099 &     Zhang, L. \\
\hline
\end{tabular}
\end{center}
\end{table}

\subsection{Transit depth as a function of wavelength}
\label{chomatic_lcs}

\begin{figure*}
   \centering
   \includegraphics[height=\hsize,angle=270]{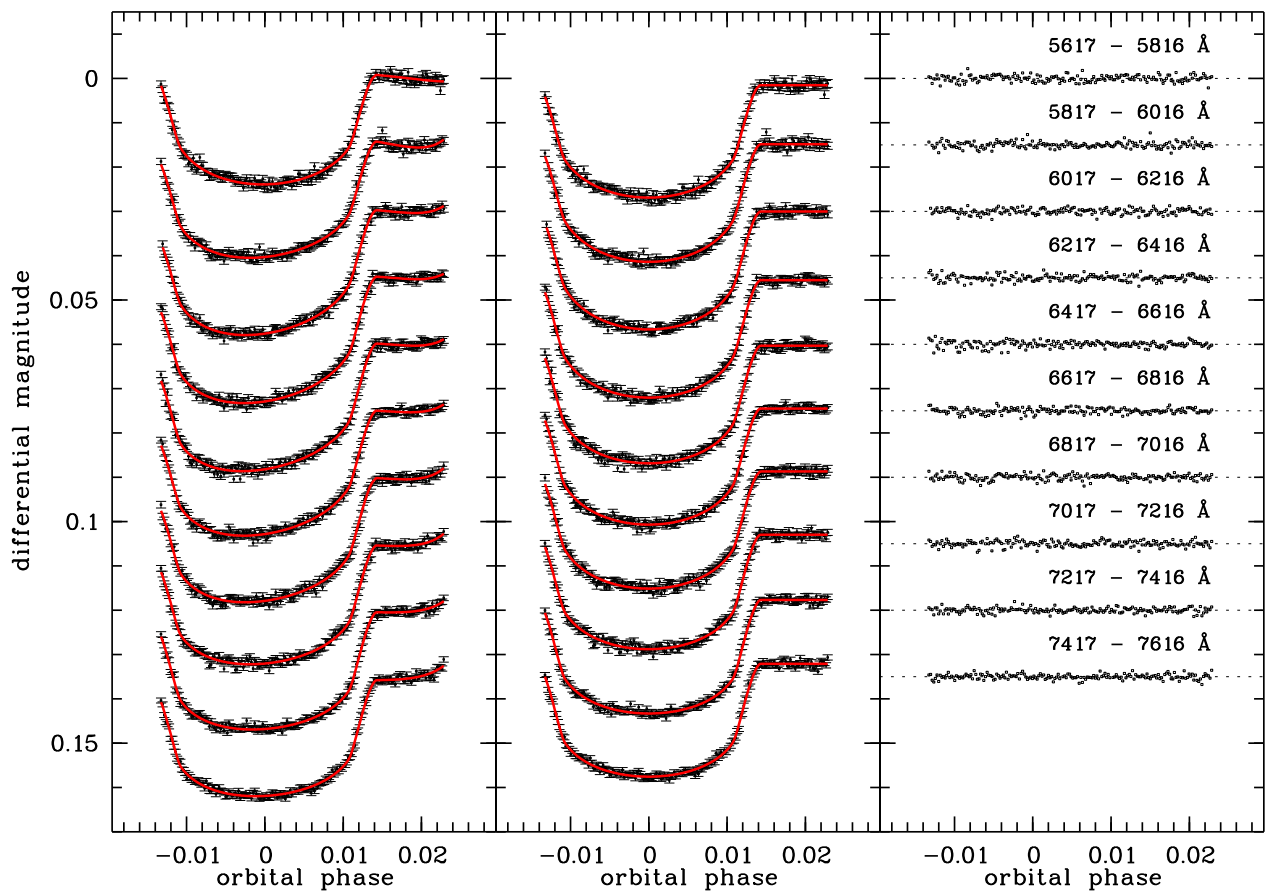}
      \caption{Chromatic set of transit light-curves of HAT-P-19b, flux integration made in 200 \AA{} wide flux channels. Left panel: the raw light curves, middle panel: detrended light curves, right panel: the light-curve residuals. }
         \label{200er_lcs}
   \end{figure*}

We created two different chromatic sets of light curves by binning the flux of our spectra in wavelength channels of a certain width. At first, we chose 200 \AA{} as channel width, which yielded rather high S/N per photometric data point, but a poor spectral resolution in our resulting transmission spectrum. Then we binned the flux in narrower channels, which provided a higher spectral resolution at the cost of noisier light curves and therefore higher uncertainties of the transit parameter. We performed a simple estimation of the channel width that was most sensitive to narrow spectral features with the help of a theoretical 1000 K transmission spectrum \citep{Fortney2010}. We compared the theoretically predicted transit depth in a channel centered on the Na D line core with the transit depths in the adjacent channels for channel widths of 10 to 150 \AA{}. The function of differential transit depths over channel width was weighted with a simple square-root law to account for the S/N 
dependence on the width. For simplicity we assumed here photon noise to dominate the noise budget. The resulting function peaks at about 50~\AA{}. Figure \ref{200er_lcs} shows the light-curve set of 200 \AA{} channel width. In the left column the raw light curves are shown, in the middle column the detrended light curves, and in the column to the right the residuals versus the detrend + transit model.

The chromatic set of light curves was analyzed simultaneously. The fitting parameters per light curve were the four coefficients of the airmass-dependent third-order polynomial, $u_{\mathrm{A}}$, and $k$. In addition, we fitted for a common value per set of $a / R_A$, $i$, and $T_0$ with the white-light curve uncertainties of Table \ref{KapSou} as Gaussian priors. In total, the number of simultaneously fitted parameters were 63 for the 10 light curves of 200 \AA{} width, and 249 for the 41 light curves of 50 \AA{} width. The results are presented in Table \ref{rmslc1}. 

We verified the MCMC results of $k$ over wavelength with another available transit-modeling package, JKTEBOP\footnote{www.astro.keele.ac.uk/jkt/codes/jktebop.html} \citep{Southworth04}. It uses the Levenberg-Marquardt optimization algorithm to find the best-fitting model and includes multiple options to give reliable error estimations. Here, we calculated the errors with a Monte Carlo simulation \citep{Southworth2005} and a residual-permutation algorithm \citep{Jenskins2002,Southworth08}, the latter being sensitive to correlated noise. We adopted the higher value of both. The vast majority of best-fit parameters agrees within one sigma, and all conclusions derived in this work about the measured transmission spectrum of HAT-P-19b are the same for both the MCMC analysis and JKTEBOP analysis. On average, the uncertainties of JKTEBOP were larger by about 20\% than the uncertainties of the MCMC chains. This deviation does not change the results of this work, and a detailed analysis of its origin is beyond the 
scope of this work.

   \begin{figure*}
   \centering
   \includegraphics[height=\hsize,angle=270]{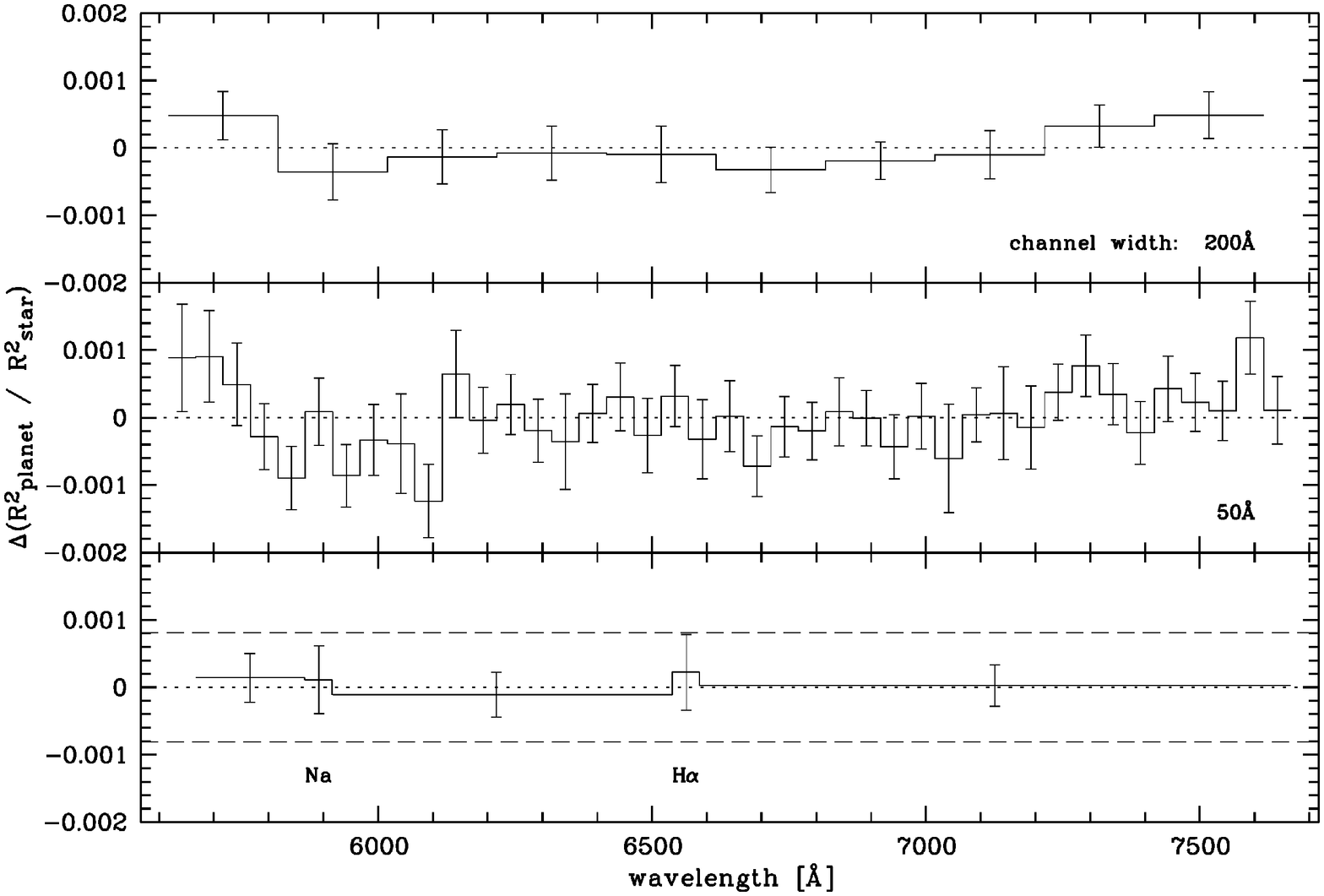}
      \caption{Transmission spectrum of HAT-P-19b. In the upper panel the relative change in transit depth is shown for 200\AA{} wide wavelength channels, in the middle panel a channel width of 50 \AA{} is presented. For the lower panel two 50 \AA{} wide channels concentrated on the line center of the Na doublet at $\sim$5892 \AA{} and H$\alpha$ at $\sim$6563 \AA{}. No significant variation was detected. The horizontal dotted line in the lower panel marks the average value, the dashed lines denote the difference corresponding to three atmospheric scale heights. }
         \label{transspec}
   \end{figure*}

The transmission spectrum plotted in Fig. \ref{transspec} appears very flat without significant outliers. We calculated the $\chi^2$ of the 50 \AA{} data set versus a Rayleigh-scattering slope given by $dR_b/d\mathrm{ln}\lambda\,=\,-4H$ similar to the observed spectrum of HD189733b \citep{Leca2008b} with $H$ as the atmospheric scale height. We also compared the data to a 1D solar metallicity model using a planet-wide averaged P-T profile computed by \cite{Fortney2010} for the parameters of the HAT-P-19 system, see Fig. \ref{transspec_model}. Both models were binned in the same 50 \AA{} wavelength channels as the observed spectra. The cloud-free model shows a prominent sodium absorption feature with pressure-broadened wings, also predicted by other modeling work for HAT-P-19b's equilibrium temperature of about 1000 K \citep{Seager2000,Brown2001,Hubbard2001}.  A similar feature of potassium causes the upward slope of planetary size in the model at the red end of our wavelength range, the line core of 
potassium was not covered, unfortunately. The lowest $\chi^2$ of 34.7 for 40 degrees of freedom is given by a flat line, 36.8 was obtained for the Rayleigh-scattering spectrum and 48.9 was the value for the cloud-free atmosphere.

We tried to test the influence of certain steps of our analysis on the obtained transmission spectrum. For example, we shifted the flux channels in wavelength by fractions of their width, we tested different lengths of post-transit baseline and the effect of fixing $a/R_A$ and $i$ to the values found by H11. We also varied our treatment of the limb-darkening coefficients by fixing both $u_{\mathrm{A}}$ and $v_{\mathrm{A}}$ to theoretical values, another time we fitted for both in the transit modeling. No test yielded a significant deviation in spectrum shape from the spectrum presented here in Fig. \ref{transspec}.

\subsection{Photometric variability of HAT-P-19}
\label{ana_lotephot}

\begin{figure}[b]
   \centering
   \includegraphics[height=\hsize,angle=270]{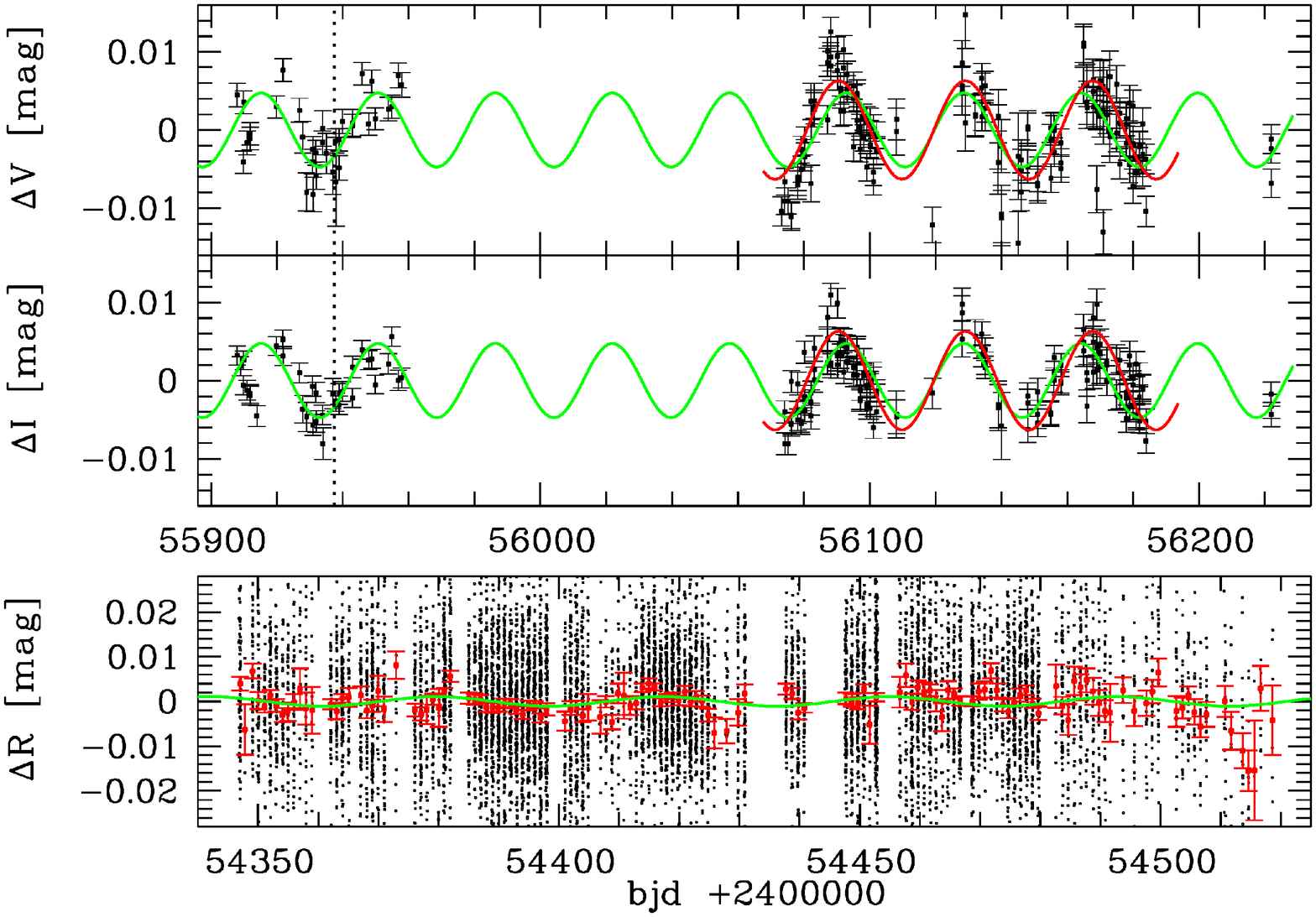}
      \caption{Upper panel: STELLA differential V band photometry. A simple sine model of the 2011+2012 data set overplotted in green, a sine model of just the 2012 data in red. The vertical dotted line marks the transit time of our spectroscopic observation. Middle panel: The same as in the upper panel, here for filter I. Lower panel: The HATnet differential R band photometry from 2007/2008. Black dots are the individual data points, red is the nightly average and green gives the best fit sine model.}
         \label{lotephot}
   \end{figure}

\begin{figure}[b]
   \centering
   \includegraphics[height=\hsize,angle=270]{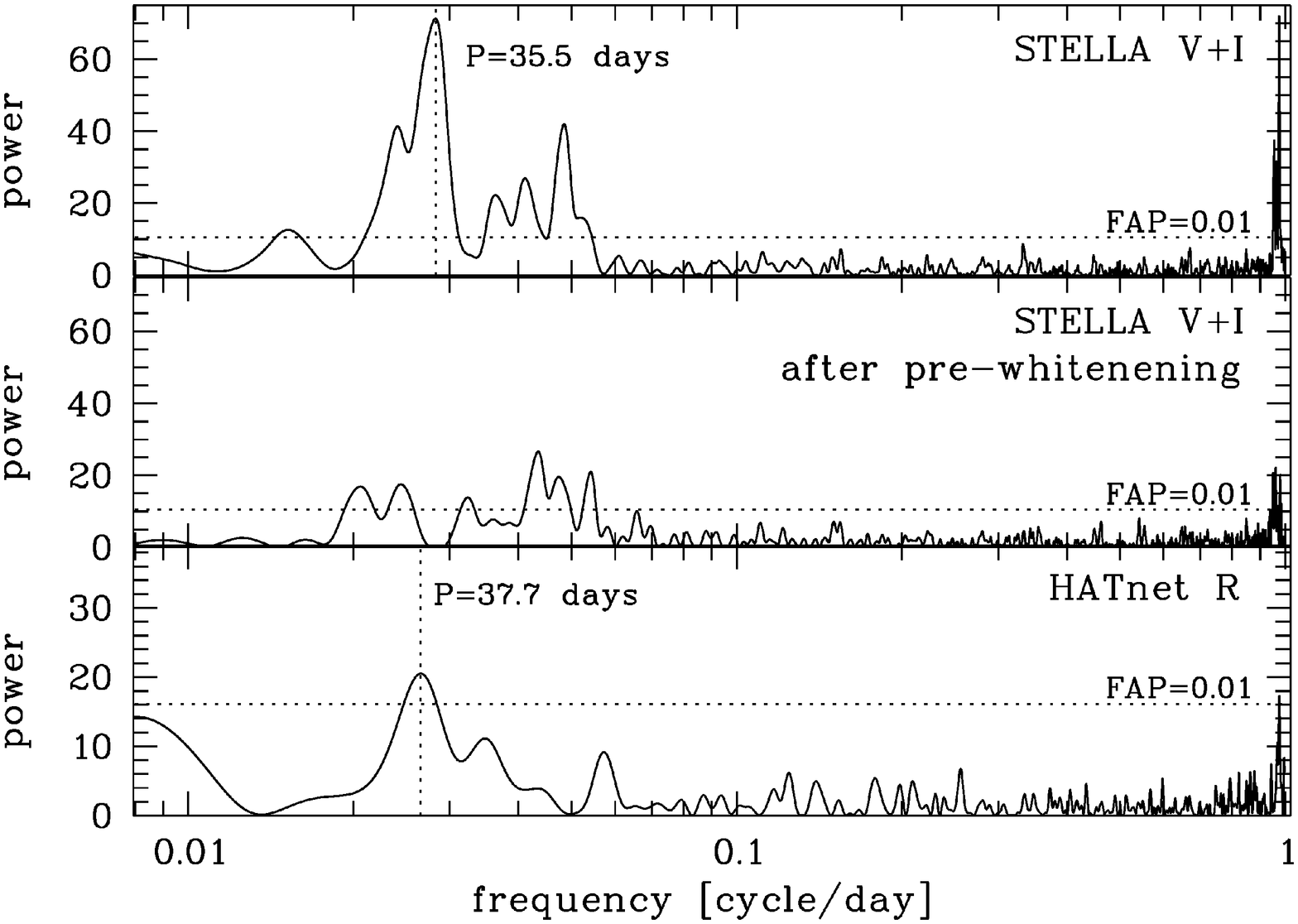}
      \caption{Lomb-Scargle periodogram for the STELLA photometry (upper panel) and the HATnet photometry (lower panel). Horizontal dashed lines denote the power value of false-alarm probability (FAP) 0.01. Vertical dashed lines mark the period found by the sine fit. The middle panel shows the STELLA periodogram after one pre-whitening step. Significant frequencies remain, which we attribute to the sampling pattern and/or spot evolution.}
         \label{periodogram}
   \end{figure}

We found a clear periodic variation of HAT-P-19 in the long-term photometry shown in Fig.\ref{lotephot}. A least-squares sine fit to the data revealed a period of $35.5 \pm 2.5$ days. The error value is drawn from the full width at half maximum of the peak in the periodogram in Fig. \ref{periodogram}. The amplitude was different for the used filters, V and I, with $4.7 \pm 0.5$ and $3.0 \pm 0.4$~mmag provided by the least-squares sine fit. There are indications of varying period and amplitude in our data. When analyzing only the data of the observing season 2012, we obtain a stronger amplitude of $6.3 \pm 0.6$ and $4.6 \pm 0.5$~mmag for V and I and a slightly longer period of $38.6 \pm 4.4$ days. In the interpretation of the flux variation that is caused by starspots rotating in and out of view, these changes in period and amplitude are possibly related to spot evolution. However, the period variation is within one $\sigma$ and the amplitude variations do not exceed 3 $\sigma$.
 
The periodogram in Fig. \ref{periodogram} shows three additional peaks next to the main signal at somewhat shorter period. They loose most of their power or disappear entirely when subtracting the most significant signal (pre-whitening). Instead, in the residual data we find another signal above the false-alarm probability of 0.01 at about 23 days. This period might also be related to spot evolution, but an in-depth period analysis is beyond scope of the current work.

H11 did not mention a stellar flux variability. We reanalyzed the HATnet photometry observed from September 2007 to February 2008 to search for the same periodicity. The available time series\footnote{http://www.hatnet.org/planets/discovery-hatlcs.html} was treated by the external parameter decorrelation technique and the trend-filtering algorithm, as is common for HATnet data (see H11 and references therein). When we calculated a Lomb-Scargle periodogram, we found a significant period of $P=37.7\pm3.1$ days (false-alarm probability FAP < 0.0002), see Fig. \ref{periodogram}. The sine fit reveals a much lower variability amplitude than the STELLA data of only $1.0 \pm 0.3$~mmag.

\subsection{Correction for starspots}
A likely reason for the periodic flux variation of HAT-P-19 are cool spots rotating in and out of view since the variation amplitude is larger in the V band than in the redder I band. Brightness inhomogeneities on the stellar surface influence the transit parameters \citep{Pont2008,Czesla2009}. An occultation of a starspot by the transiting planet causes a bump in the light curve, which leads to a shallower transit and an underestimation of the planet size if uncorrected. Starspots unocculted from the planet cause the average brightness to be higher along the transit chord than on the rest of the stellar hemisphere, which leads to deeper transits and an overestimation of the planet size. Examples of active host stars with frequently observed starspot crossings are HD189733 (see \citealt{Pont2013} and references therein) and WASP-19 \citep{Tregloan2013,Huitson2013}. 

The white-light curve gives no evidence of a starspot crossing event. Possible scenarios that allow for spots along the chord without resolvable deformations of the transit shape such as very small spots or a homogeneous spot distribution along the chord are considered to either be negligible in their effect or unlikely. We therefore assume that the transit depth is not underestimated. With the long-term photometry on hand, we can estimate the influence of unocculted starspots on the stellar hemisphere visible at the moment of the transit event. This estimation depends on three quantities: The stellar flux level $f_{\mathrm{meas}}$ at the time of the transit observation, the stellar flux level without spot dimming $f_{\mathrm{quiet}}$ , and the effective temperature of the spots $T_{\mathrm{spot}}$. We define the relative variation in flux $\Delta f$ by 
\begin{equation}
\Delta f\,=\,\frac{f_{\mathrm{meas}}-f_{\mathrm{quiet}}}{f_{\mathrm{quiet}}}
,\end{equation}
with $f_{\mathrm{meas}}$ as the measured stellar flux and $f_{\mathrm{quiet}}$ as the flux of the quiet star without brightness inhomogeneities. \cite{Desert2011} described how $\Delta f $ simply translates into the change in transit depth by
 \begin{equation}
\frac{k^2_{\mathrm{meas}} - k^2_{\mathrm{true}}}{k^2_{\mathrm{true}}}\,=\,\alpha \Delta f.
\label{delta_k}
\end{equation}
The proportionality factor $\alpha$ was estimated to be $-3$ in the system HD189733 \citep{Desert2011}. Other studies simplified Eq. \ref{delta_k} for other planetary systems by $\alpha=-1$ (e.g., \citealt{Berta2011}). We cannot measure this on our own without transit measurements at multiple epochs, therefore we followed the simplest approach and set $\alpha$ to -1.

A value of the flux variation $\Delta f$ includes the knowledge of HAT-P-19's flux level $f_{\mathrm{quiet}}$ without any spots on its visible surface. Spots were probably also present at the time of the highest count rate of our light curve, therefore $f_{\mathrm{quiet}}$ needed to be estimated. We approximated the level of a permanent flux dimming by the level of the variance of the long-term light curve, an approach undertaken by \cite{Pont2013} for HD189733 based on the work of \cite{Aigrain2012}. Thus, at the time of our transit observation we estimate $\Delta f \sim -0.007$ in V band. Conservatively, we rounded this value to 1\% flux dimming by spots.

To estimate $\Delta f$ over the spectral range covered by our transit observation, a value of the starspot temperature is needed.  However, $T_{\mathrm{spot}}$ cannot easily be derived by photometric data because of a degeneracy between spot size and spot temperature \citep{Strassmeier2009}. In principle, this can be broken by multicolor photometry, but our error bars on the subpercent variation amplitudes in V and I band cannot reasonably constrain the temperature contrast between spots and photosphere. The starspots of the exoplanet host star HD189733, similar in stellar parameters to HAT-P-19, have temperatures of between 750 K \citep{Sing2011b} and 1000 K \citep{Pont2008} lower than the spot-free photosphere. \cite{Berdyugina2005} listed several early-K dwarfs with spots of about 1500~K temperature contrast. We calculated spot corrections on the transmission spectrum for three different temperature contrasts, 500~K, 1000~K, and 1500~K. The spectral energy distributions of the spots and 
spot-free photosphere were approximated by blackbody radiation. Furthermore, we neglected the contribution of faculae. Under these assumptions, a flux dimming of one percent corresponds to about 2.3 (1.4, 1.1) percent coverage of the stellar hemisphere with spots of 500~K (1000~K, 1500~K) temperature contrast. The obtained $\Delta k^2$ correction values over wavelength range from  $\sim$1\% of the transit depth ($\Delta k^2 \sim 0.00019$) at the blue end of our spectrum to about 0.8\% ($\Delta k^2 \sim 0.00016$) at the red end. The difference in $\Delta k^2$ according to the different spot temperatures is at the level of 0.1\% of
the transit depth at 7500 \AA{}. Our derived uncertainty of the transit depth $k^2$ for the white-light curve is about 1.7\%. Hence, the spot correction on the absolute transit depth is of minor importance, and the correction on the relative changes of transit depth over wavelength is negligible.

\section{Discussion}

\subsection{Transmission spectrum}

   \begin{figure}
   \centering
   \includegraphics[height=\hsize,angle=270]{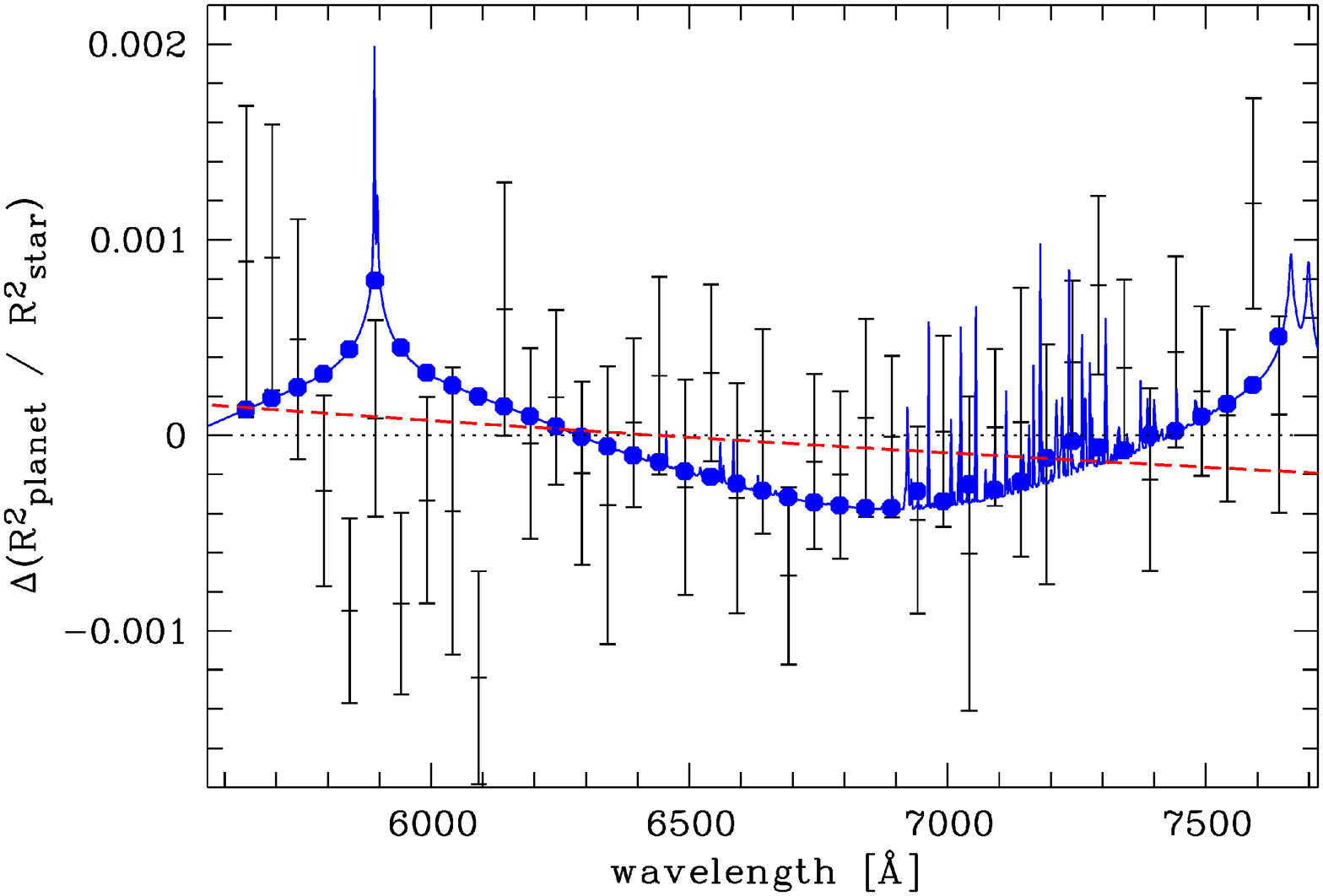}
      \caption{Derived transmission spectrum in comparison to a theoretical spectrum calculated by \cite{Fortney2010} illustrated as a continuous blue line. The model values integrated over 50 \AA{} flux channel width are shown as blue filled circles. We
also present a spectrum typical for Rayleigh-scattering as a
red dashed line. Both models were computed according to the system parameters of HAT-P-19. }
         \label{transspec_model}
   \end{figure}
 
We ran a simple calculation to test whether any potential variations in transit depth among the narrow wavelength channels might be blurred by the point spread function. We assumed the FWHM in dispersion direction to equal the measured FWHM in spatial direction due to the 5'' wide slit. While the FWHM increased to more than ten pixel toward our OoT phase limit of 0.0228, the average value during second and third contact of the transit was 4.8 pixel (1.2'' on the sky). The corresponding light leakage weakens a potential difference in transit depth of the sodium 50 \AA{} channel with respect to the adjacent channels by $\sim$3\%. Hence, we neglect this effect here.

Key values for an initial characterization of an exoplanet atmosphere are the error bars of the relative transit depth measurements. We found the average error value to be higher than the scatter of the $k^2$ values for both the 50 \AA{} and the 200 \AA{} light-curve sets. The reduced $\chi^2$ values of 0.9 and 0.7, respectively, versus a constant $k^2$ confirm this indicator of a slight overestimation of the error bars. We compared our achieved 200 \AA{} accuracy to similar investigations present in the literature by a search for values of ground-based single-transit analyses using a comparable width of the flux channels in about the same wavelength range. We caution that differences can occur due to different telescope size, target brightness, and the complexity of systematics in the light curves and analysis techniques. However, our average uncertainty of $k^2$ for the 200 \AA{} channels of $3.6\cdot10^{-4}$  agrees well with the range of literature values $1.8 - 3.4\cdot10^{-4}
$ \citep{
gibson2013,GibsonHATP32,Jordan2013,Murgas2014}. One reason for our uncertainty to be slightly larger is certainly the lack of an out-of-transit baseline on one side.

Our transmission spectrum of HAT-P-19b favors a featureless flat spectrum, but although less likely, the Na rich, cloud-free atmosphere overplotted in Fig. \ref{transspec_model} is not ruled out. If confirmed by follow-up observations, a non-detection of the pressure-broadened Na absorption could be explained in two ways: First, a cloud or haze layer does not allow probing the deeper layers of the atmosphere and covers the broad alkali line wings. As a result
of the slanted viewing geometry during transmission spectroscopy, even condensates of rather low abundance can make the atmosphere opaque \citep{Fortney2005}. Observations of higher spectral resolution are needed to search for a spectroscopic feature at the line core. The other explanation would be a clear atmosphere with a depletion of atomic sodium. Sodium condensation into clouds of sodium sulfide, for instance, at the planets cooler night side could potentially decrease the amount of atomic sodium at the terminator region probed by our measurement \citep{
Lodders1999}. Another potential cause of atomic Na depletion is ionization by stellar UV photons \citep{Fortney2003}, a scenario which is plausible because we found HAT-P-19 to be magnetically active.  

Among the known sample of hot-Jupiter exoplanets that have already been investigated by transmission spectroscopy there are three planets similar to HAT-P-19b in their system parameters: HAT-P-1b \citep{Bakos2005}, HAT-P-12b \citep{Hartman2009}, and WASP-29b \citep{Hellier2010}. All these planets have roughly the mass of Saturn, roughly the size of Jupiter, and an equilibrium temperature of 1000 to 1200~K. They orbit solar-like main-sequence stars of spectral type G or K. \cite{gibson2013} obtained an optical transmission spectrum of WASP-29b around the sodium line and was able to significantly rule out a pressure-broadened absorption feature. \cite{Nikolov2013} observed HAT-P-1b in the optical regime during transit and detected additional absorption at the sodium core in a 30 \AA{} wide flux channel, but no spectral signature of the broadened line wings. \cite{Line2013} observed HAT-P-12 in the near-IR and found evidence against spectral signatures of H$_2$O tha tis theoretically expected in a clear 
atmosphere. 
All these studies have in common that they disagree with theoretical cloud-free atmosphere models, but could be much better explained by an opaque layer in the atmosphere either blocking only the pressure-broadened wings or the entire absorption feature when observed in low spectral resolution. Our measurements for HAT-P-19b agree well with these results and point toward the final statement of \cite{Fortney2005}, who reported that owing to the slanted viewing geometry of transmission spectroscopy, the detection of clear atmosphere will be rare. The only object for which sodium wings, which indicate a clear atmosphere, have been detected so far is the somewhat hotter HD209458b \citep{Sing2008a}. But also in the temperature regime of 2000~K and above, it seems to be very common to find a spectrum more depleted in features than expected by clear atmosphere models, see the review by \cite{Madhusudhan2014}.

Future follow-up observations of HAT-P-19b with HST/WFC3 from 1.1 to 1.7~$\mu$m could strengthen our present indication for an atmosphere containing clouds or hazes by investigating the water absorption band at 1.4~$\mu$m.

\subsection{Long-term photometric variability}

H11 gathered photometry data with the HATnet telescopes from September 2007 to February 2008, high-resolution spectra for RV follow-up have been taken in August 2009 to August 2010. The
authors did not mention any indication for stellar activity from photometry or spectroscopy. The RV jitter is low, 6.7 m/s, and might be caused by instrumental systematics. However, the periodic photometric variations found by us in data taken from December 2011 to Octobre 2012 were interpreted as starspots moving in and out of view because of the larger amplitude of the variation in B than in V. The same period of $\sim$37 days detected in the HATnet data and STELLA data, consequently interpreted as stellar rotation period, shows an increase in amplitude from 2007 to 2012 by about a factor of 5. Possible interpretations for this increase are a very homogeneous spot distribution in 2007 and a very inhomogeneous distribution in 2012. Another explanation is a change in activity level comparable with changes in the activity level during the 11-year solar cycle. 
Such stellar activity cycles have been detected in a variety of stars, see \cite{Strassmeier2009} and references therein. 

Knowing the rotation period of the star and its effective temperature, we have the opportunity to determine the age of HAT-P-19 using gyrochronology. Here, we assumed that the host star was not spun up by tidal interactions with its close-in planet. According to Eq. 15 in \cite{Husnoo2012}, the timescale for synchronization of the rotation is above the age of the Universe and several orders of magnitudes higher than for systems with detected excess rotation that might be caused by tidal effects \citep{Pont2009,Husnoo2012,Poppenhaeger2014}. To apply the age determination from \cite{Barnes2010}, the effective temperature was converted empirically to a B-V color by using the formula in \cite{Reed1998}. We derived $\mathrm{B}-\mathrm{V}=0.927\pm0.042$ from T$_\mathrm{eff} = 4990 \pm 130$~K. The errors arising from the unknown initial rotation period at zero age of a star between 0.12~d and 3.4~d result in an uncertainty of 0.2~Gyr for that specific spectral type and rotation period. Taking all 
uncertainties into account (error in effective temperature, error in period determination, error in age determination), we estimate the age of the exoplanet host HAT-P-19 to be $5.5^{+1.8}_{-1.3}$ Gyr. This result refines the age estimation from H11 of $8.8 \pm 5.2$ Gyr based on stellar evolutionary tracks.

Knowledge of the rotation period of the star and its spectroscopic line broadening $vsini$  in principle allows determining the inclination of the star in the observer's line of sight \citep{Walkowicz2013}. It does not provide an absolute angle between the orbital plane of the planet and its host star rotation axis, but as we know the inclination of the planetary orbital plane from the transit measurement to be close to 90\textdegree, any significant deviation of the stellar inclination from the same value indicates a misalignment of the planetary system. However, for very slowly rotating stars like HAT-P-19, no meaningful conclusions can be drawn. The $vsini$ value of $0.7 \pm 0.5$ km/s (taken from H11) in combination with our stellar rotation period of $\sim$35 days allows for inclination angles from 10 to 90 degrees.

\section{Conclusion}

We reported GTC/OSIRIS spectroscopic observations of one transit of HAT-P-19b covering the wavelength range from 5600 to 7600 \AA{}. The data allowed us to refine the transit parameters and the planetary orbital ephemeris. We derived an optical transmission spectrum of the Saturn-mass highly inflated exoplanet including the wavelength range of the Na D line. We extracted a transmission spectrum using 200 \AA{} wide wavelength channels (spectral resolution $R\sim30$) and another using 50 \AA{} wide channels ($R\sim120$). The result is in both cases a flat spectrum without additional absorption at any specific wavelength and without significant trends. No sodium absorption feature of the planetary atmosphere was detected.

We performed a monitoring campaign to search for a photometric variability of HAT-P-19 using the robotic 1.2m telescope STELLA in the years 2011 and 2012. We found a clear periodic variation with $4.7 \pm 0.5$ mmag and $3.0 \pm 0.4$ mmag amplitude in the filters V and I and a period of $\sim$37 days. The same periodicity of lower amplitude was found in the HATnet discovery photometry of 2007 and interpreted by us as the stellar rotation period. Our monitoring data covered the date of the spectroscopic transit observation and could therefore be used to correct the transit parameters for systematics caused by spots on the visible hemisphere. However, the influence of the spots on the transmission spectrum was found to be negligible. The obtained rotation period of the host star enabled us to refine the age estimate of the star by gyrochronology.

The derived transmission spectrum of HAT-P-19b favors a featureless gray atmosphere, but comparisons with theoretical models tuned to the HAT-P-19 system parameters showed that the achieved accuracy is not sufficient to place further constraints on the atmospheric composition. A model showing a Rayleigh-scattering slope and a solar-composition cloud-free atmosphere model with a pressure-broadened sodium line roughly agrees with the data. A transmission spectrum of higher S/N ratio can be obtained easily by follow-up observations of the same kind as presented here, since the current work is based on a single-transit observation in imperfect observing conditions. The lack of pre-ingress baseline made our modeling of underlying systematics in the light curve (detrending) more error prone and hence increased our final error bars in the transmission spectrum. If follow-up observations confirm the flat transmission spectrum to higher significance, the spectrum of HAT-P-19b would share an 
important property with the spectra of other close-in gas giants of about 1000 K atmospheric temperature investigated so far \citep{gibson2013,Nikolov2013,Line2013}: None of their transmission spectra can be explained by a cloud-
or haze-free atmosphere model, they all show indications of an additional opacity source blocking parts of the probable atmosphere.
%

\begin{acknowledgements}
We thank the entire GTC staff for their help in executing this program. We thank Jonathan Fortney for providing the cloud-free solar composition atmospheric model for HAT-P-19b. We thank Joel Hartman for providing the HATnet photometry data. I. R. and E. H. acknowledge financial support from the Spanish Ministry of Economy and Competitiveness (MINECO) and the ``Fondo Europeo de Desasarrollo Regional'' (FEDER) through grant AYA2012-39612-C03-01. This work made use of PyAstronomy.
\end{acknowledgements}

\bibliographystyle{aa}
\bibliography{mybib}

\end{document}